\documentclass[pre,showpacs]{revtex4}

\begin{document}

\title{Investigation of a generalized Obukhov Model for
Turbulence}

\author{A. Baule and R. Friedrich}

\affiliation{Institute of Theoretical Physics,
Westf\"alische Wilhelms-Universit\"at M\"unster,
Wilhelm-Klemm-Str. 9, 48149 M\"unster, Germany}

\begin{abstract}

We introduce a generalization of Obukhov's model
[A.M. Obukhov, Adv. Geophys. 6, 113 (1959)]
for the description of the 
joint position-velocity statistics of a single fluid particle in fully
developed turbulence. In the presented
model the velocity 
is assumed to undergo a continuous 
time random walk. This takes
into account long time correlations. As a consequence 
the evolution equation for the joint position-velocity probability 
distribution
is a Fokker-Planck equation with a fractional time derivative.
We determine the solution of this equation
in the form of an integral transform and derive
a relation for arbitrary single time moments. Analytical solutions
for the joint probability distribution and its moments are given.

\end{abstract}

\maketitle

\section{Introduction}

The statistics of turbulent flows is described by phenomenological theories
dating back to Kolmogorov and Obukhov \cite{Monin,Frisch,Pope}.
In a Lagrangian treatment the path 
${\bf X}(t,{\bf y})$ and the velocity ${\bf U}(t,{\bf y})$ of 
advected particles initially starting at the position ${\bf y}$ are determined
by the acceleration
${\bf A}(t,{\bf y})$ which in principle
is given by the Navier-Stokes equation: 
\begin{eqnarray}\label{accel}
\frac{d{\bf X}(t,{\bf y})}{dt}={\bf U}(t,{\bf y})\qquad,\nonumber\\
\frac{d{\bf
U}(t,{\bf y})}{dt}={\bf A}(t,{\bf y})
\qquad.
\end{eqnarray}
The central statistical quantity 
is the joint position-velocity probability distribution of the
particle: 
\begin{eqnarray}\label{pdf} f({\bf u},{\bf x},t)=\:<\delta({\bf
x}-({\bf X}(t,{\bf y})-{\bf U}(0,{\bf y})t-
{\bf y}))\delta({\bf u}-({\bf U}(t,{\bf y})-
{\bf U}(0,{\bf y}))>\qquad, 
\end{eqnarray}
where the brackets denote a suitable average over a stationary statistical
ensemble. Here and in the following
we shall be interested in the statistics of increments 
${\bf u}(t)={\bf U}(t,{\bf y})-{\bf U}(0,{\bf y})$,
${\bf x}(t)={\bf X}(t,{\bf y})-{\bf U}(0,t)t-{\bf y}$. 
The pdf (\ref{pdf}) obeys the 
initial condition, $f({\bf u},{\bf x},0)=\delta({\bf u})\delta({\bf x})$.
Furthermore, Kolmogorov's theory of 1941  
(K41) \cite{Monin} suggests the existence of scaling behaviour
\begin{equation}\label{norm}
f({\bf u},{\bf x},t)=\frac{1}{t^6} f_0(\frac{{\bf u}}{\sqrt{t}},
\frac{{\bf x}}{\sqrt{t}^3}) \qquad .
\end{equation}
Although such a scaling behaviour may not exist due to 
intermittency \cite{Frisch} the assumption of
normal scaling behaviour (\ref{norm}) usually 
serves as a first starting point.   
\\
Due to the fact that a successfull statistical approach based on a treatment
of the Navier-Stokes equation is still missing, one has to resort to
heuristic assumptions about the statistics of ${\bf A}(t,{\bf y})$.
An early model for $f({\bf u},{\bf x},t)$ has been introduced
by Obukhov \cite{Obukhov}, who assumed, that the turbulent 
acceleration ${\bf A}(t,{\bf y})$ is a Gaussian, 
$\delta$-correlated random force:
\begin{eqnarray}
<A_i(t,{\bf y})A_j(t',{\bf y})>\:=2\delta_{ij}\delta(t-t') \qquad.
\end{eqnarray}
In turn the probability distribution (\ref{pdf}) 
obeys a Fokker-Planck equation:
\begin{eqnarray}\label{7-Ob_FPE}
\frac{\partial}{\partial t}f({\bf u},{\bf x},t)
+{\bf u}\cdot\nabla_{{\bf x}} f({\bf u},{\bf x},t)
=\Delta_{{\bf u}} f({\bf u},{\bf x},t)
\qquad.\end{eqnarray}

The solution of the Obukhov-model with the initial condition
$f({\bf u},{\bf x},0)=\delta({\bf x})\delta({\bf u})$ possess
a Gaussian form exhibiting scaling behaviour
\begin{eqnarray}
<{\bf u}^2(t)>\sim t\qquad,\qquad <{\bf x}^2(t)>\sim t^3 \qquad,
\end{eqnarray}
which is consistent with the phenomenological scaling theory of Kolmogorov
(K41). However Obukhov's model contradicts recent experimental results
 \cite{Pinton}, \cite{Boden}: 
the probability distribution for the velocity increment is far 
from Gaussian and the
increment behaviour can not be a simple random walk as suggested by Obukhov. 
The reason is the intermittent character of turbulent flows. 
Nonnormal statistics for Lagrangian variables 
may originate from long time correlations in the turbulent field.
During the last few years lots
of efforts have been put into the formulation of more sophisticated 
phenomenological theories which
can take into account these facts (c.f. the review article \cite{Aringazin}).
Recently a connection
between the velocity increment statistics of a Lagrangian particle 
and a type of continuous time
random walk has been introduced \cite{Friedrich} by a truncation of
an infinite chain of evolution equations for multiple particle probability
distributions. The following 
evolution equation for the single time joint position velocity increment
probability distribution has been obtained 
\begin{eqnarray}\label{frimo}
\lbrace
\frac{\partial }{\partial t}+{\bf u}\cdot \nabla_x \rbrace
f({\bf u},{\bf x},t)&=& \frac{1}{\Gamma(\alpha)} \frac{\partial }{\partial t} \int_0^t 
\frac{dt'}{(t-t')^{1-\alpha}} \nabla_u Q({\bf u}^2) 
[\nabla_u f({\bf u},{\bf x}-{\bf u}(t-t'),t')]_{{\bf u}'={\bf u}} \qquad,
\nonumber \\
Q_{ij}({\bf u}^2)& =& {\bf u}^{2(1-\alpha)}[\delta_{ij}Q_1+Q_2 \frac{u_iu_j}{
{\bf u}^2}] \qquad .
\end{eqnarray}
($\Gamma(\alpha)$ denotes the Gamma- function). In the case of 
isotropic turbulent flows, the  
matrix $Q_{ij}({\bf u}^2)$ has to be invariant
with respect to rotations leading to the form given above. 
Furthermore, the variables $Q_1$ and $Q_2$
are constant due to the fact that solutions of the above
equation are required to allow for 
scaling behaviour of the velocity increment
$u \approx t^{1/2}$ (for constant value of $\alpha$).
The parameter $\alpha$ is taken from the interval $0 < \alpha \le 1$.
This equation generalizes Obukhov's model in several respects. 
First, it introduces a temporal memory. Second, the simple diffusion process 
in velocity space is changed to a diffusion process
with velocity dependent diffusion coefficient. Third, retardation 
effects with respect to the spatial coordinate
appear. It has been shown
that solutions of the resulting equation for 
the velocity pdf $G({\bf u},t)=\int d{\bf x}f({\bf u},{\bf x},t)$ yields 
excellent approximations to the experimentally determined velocity pdf's,
provided the parameter $\alpha$ is allowed to vary with the time increment.
A complete solution to the equation (\ref{frimo}) has not yet been 
obtained.   

In the present paper we investigate a simpler 
phenomenological model by disregarding the retardation in the spatial
coordinate:
\begin{eqnarray}\label{model}
\lbrace
\frac{\partial }{\partial t}+{\bf u}\cdot \nabla_x \rbrace
f({\bf u},{\bf x},t)= 
{_0D}_t^{1-\alpha} L({\bf u},\nabla_u)
f({\bf u},{\bf x},t) +\delta(t) \delta({\bf x})\delta({\bf u}) \qquad.
\end{eqnarray}
Here, $L({\bf u},\nabla_u)$ is a diffusion operator with a velocity dependent
diffusion constant (compare eq. (\ref{frimo})). We have explicitly included
the initial condition.
The operator
$_0D_t^{1-\alpha}$ denotes the Riemann-Liouville fractional differential
operator (see e.g. \cite{Podlubny}): 
\begin{eqnarray}
\label{FracDiff_def}
_0D_t^{1-\alpha}F(t)=
\frac{1}{\Gamma(\alpha)}\frac{\partial}{\partial t}\int_0^t
\frac{F(t')}{(t-t')^{1-\alpha}}\:dt' \qquad.
\end{eqnarray}
This is a straightforward generalization of integer order 
differentiation to fractional orders.
The presence of the
fractional differential operator eq.(\ref{FracDiff_def}) introduces 
temporal memory effects depending on the parameter $\alpha$. We
assume that $0<\alpha \le 1$. In the limit $\alpha \rightarrow 1$
 our model eq.(\ref{model}) reduces to the
ordinary Obukhov model eq.(\ref{7-Ob_FPE}) due to the
property $_0D_t^{0}F(t)=F(t)$ if one takes $L({\bf u},\nabla_u)$ to be
the diffusion operator $L=\Delta_u$.

The purpose of the present paper is not to state the accurate evolution 
equation for the Lagrangian position-velocity increment. This task has to
be performed on the basis of a theoretical
analysis of the Navier-Stokes equation, as
has been started in \cite{Friedrich}, or by sophisticated data analysis
of the experimentally obtained Lagrangian path's of particles.
An assessment of the underlying stochastic process, in our opinion, has to 
include the description of multiple time distributions. Such an
analysis for a class of continuous time random walks has recently been
started \cite{Baule}. The purpose of the present paper is
to consider an extension of Obukhov's model to a class of fractional 
diffusion equations, whose study quite seemingly is interesting by 
its own. It is
hoped that the obtained results contribute to a more detailed 
understanding of Lagrangian turbulence statistics.   

The paper is outlined as follows. In the next section we focus on the
probability distribution for the velocity increment and review how
it can be determined by methods developed in the theory
of continuous time random walks (for a recent review of the
topic we refer the reader to \cite{Metzeler1}, \cite{Metzeler2}). 
In section III we 
shall show that the joint position velocity pdf can be
expressed as an integral transform 
\begin{equation}\label{inte}
f({\bf u},{\bf x},t)=\int d{\bf z} \int_0^\infty ds h({\bf z},{\bf x};s,t)
f_1({\bf u},{\bf z},s) \qquad ,
\end{equation}
where $f_1({\bf u},{\bf x},s)$ is the solution of 
eq. (\ref{model}) for $\alpha=1$ and 
$h({\bf z},{\bf x};s,t)$ a positive function. 
Special emphasis is put onto the case of $L({\bf u},\nabla_u)=\Delta_u$.
Here, the statistics of the velocity is characterized by 
subdiffusive behaviour related to
temporal memory effects. It is shown that both velocity and position of
the particle reveal anomalous diffusive properties. We explicitly determine
the probability distributions of velocity and spatial coordinate.   
For a diffusion operator $L({\bf u},\nabla_u)$ which allows for 
scaling solutions $f({\bf u},{\bf x},s)$ we discuss how normal scaling 
behaviour $<{\bf u}^2(t)> \sim t$ 
may arise in connection with nonnormal statistics.  

\section{The probability distribution of the velocity increment}

After integration over the position variable
in eq. (\ref{model}) an evolution equation for the marginal
probability distribution of the velocity $G({\bf u},t)=\int d{\bf x}
f({\bf u},{\bf x},t)$ is obtained in the form of the fractional diffusion
equation: 
\begin{eqnarray}\label{FDE}
\frac{\partial}{\partial t}G({\bf u},t)=\:_0D_t^{1-\alpha}L({\bf u},\nabla_u) 
G({\bf u},t)+\delta(t)\delta({\bf u})
\qquad.
\end{eqnarray}
If we take $L({\bf u},\nabla_u)=\Delta_u$ we obtain the fractional 
diffusion equation for a 
continuous time random walk \cite{Metzeler1}, \cite{Metzeler2}. 
Accordingly the second order moment reveals
subdiffusive characteristics: $<u^2(t)>\sim t^\alpha$. The corresponding
pdf is well-known and will be given below.

The solution of the
fractional diffusion equation (\ref{FDE}) 
is conveniently expressed as an integral
transform  \cite{Metzler}, \cite{Barkai}  
\begin{eqnarray}\label{FDE_sol} G({\bf
u},t)=\int_0^\infty ds\:h_\alpha(s,t)G_1({\bf u},s) \qquad, 
\end{eqnarray} 
in which the
function $G_1({\bf u},s)$ is obtained as solution of the ordinary diffusion
equation (\ref{FDE}) with $\alpha=1$ 
and the integral kernel $h_\alpha(s,t)$ is given as the single time
probability distribution of an inverse L\'evy-stable process \cite{Metzler},
\cite{Barkai}: 
\begin{equation}
\label{Barkai_n}
h_\alpha(s,t)=\frac{1}{\alpha} \frac{t}{s^{1+1/\alpha}}
L_\alpha\left(\frac{t}{s^{1/\alpha}}\right) \qquad.
\end{equation}
Here $L_\alpha(t)$ denotes a one-sided L\'evy-stable distribution of order
$\alpha$. This may be shown by inserting the ansatz (\ref{FDE_sol}) into
eq. (\ref{FDE}) leading to the solvability conditions
\begin{eqnarray}\label{hdefine}
\frac{\partial }{\partial t}h_\alpha(s,t) &=& -_0D_t^{1-\alpha} 
\frac{\partial }{\partial s}h_\alpha(s,t)  \qquad ,
\nonumber \\
_0D_t^{1-\alpha}h_\alpha(0,t) &=& \delta(t) \qquad .
\end{eqnarray}
These two equations determine $h_\alpha(s,t)$ in the form eq.(\ref{Barkai_n}).
The properties of the diffusion operator $L({\bf u},\nabla_u)$ enter 
via the pdf $G_1({\bf u},t)$. The form of the pdf, eq. (\ref{FDE_sol}) 
has the following interpretation. There is a usual diffusion process
${\bf w}(s)$ described by a Fokker-Planck operator $L({\bf w},\nabla_w)$
with respect to an {\em intrinsic time s}.
Additionally, there is a prozess $s=s(t)$ relating physical time t
and intrinsic time s with a probability distribution $h_{\alpha}(s,t)$. 
The random process ${\bf u}(t)$ is given by ${\bf u}(t)={\bf w}(s(t))$.

\section{Solution of the generalized Obukhov-model}

Motivated by the results for the fractional diffusion equation we look
for solutions of eq.(\ref{model}) which can be expressed as an integral
transform similar to eq.(\ref{FDE_sol}):
\begin{eqnarray}\label{7-Ob_LSG}
f({\bf u},{\bf x},t)=\int_{-\infty}^\infty d{\bf z} 
\int_0^\infty ds\: h({\bf z},{\bf x};s,t)f_1({\bf u},{\bf z},s)
\qquad.
\end{eqnarray}
$f_1({\bf u},{\bf z},s)$ denotes the solution of the ordinary
Obukhov-model eq.(\ref{model}) specified by $\alpha=1$:
\begin{eqnarray}\label{7-Ob_FPE2}
\frac{\partial}{\partial s}f_1({\bf u},{\bf x},s)
+{\bf u}\cdot\nabla_{{\bf x}} f_1({\bf u},{\bf x},s)
&=&L({\bf u},\nabla_u) f_1({\bf u},{\bf x},s)
\qquad.
\end{eqnarray}
The structure of the pdf, eq. (\ref{7-Ob_LSG}), has the following 
interpretation. Let us consider the case $L=\Delta_u$.
There is a stochastic process with respect to the {\em intrinsic 
time s} given by 
\begin{eqnarray}
\frac{d {\bf z}}{ds} &=& {\bf u}(s) \qquad ,
\nonumber \\
\frac{d {\bf w}}{ds} &=& {\bf F}({\bf w},s) \qquad .
\end{eqnarray}
Furthermore, there is a process $s(t)$ relating physical time t and
intrinsic time s. The velocity  ${\bf u}(t)$, thereby, is given by 
${\bf u}(t)= {\bf w}(s(t))$. If the space variable ${\bf x}(t)$ would
be obtained via ${\bf x}(t)={\bf z}(s(t))$ the function $h$ would not
depend on ${\bf x}$ and ${\bf z}$ and would be simply given by
the function $h_\alpha(s,t)$ of the preceding section.  
However, the space variable ${\bf x}(t)$
is defined by the relationship
\begin{equation}
\frac{d {\bf x}(t)}{dt}={\bf u}(t) \qquad .
\end{equation}
This explains the fact why $h({\bf z},{\bf x};s,t)$
depends on the two spatial
variables ${\bf x}$, ${\bf z}$ but not on the velocity ${\bf u}$. 

In the following we shall derive a fractional differential equation 
for $h({\bf z},{\bf x};s,t)$, corresponding boundary conditions
and obtain an explicit solution. 
We proceed as follows: the ansatz eq.(\ref{7-Ob_LSG}) is substituted into
the evolution equation (\ref{model}). The operator $L({\bf u},\nabla_u)$
only acts on $f_1({\bf u},{\bf z},s)$ such that eq.(\ref{7-Ob_FPE2}) 
can be applied:
\begin{eqnarray} \label{7-Ob_FPEsubs}
&&\int_{-\infty}^\infty d{\bf z} \int_0^\infty ds\:
\left(\frac{\partial}{\partial t}+{\bf u}\cdot\nabla_{{\bf x}}\right)h({\bf
z},{\bf x};s,t)f_1({\bf u},{\bf z},s) \nonumber \\
&&=\:_0D_t^{1-\alpha}\int_{-\infty}^\infty d{\bf z} \int_0^\infty ds\:h({\bf
z},{\bf x};s,t) \left(\frac{\partial}{\partial s}+{\bf u}\cdot\nabla_{\bf
z}\right) f_1({\bf u},{\bf z},s)
+\delta(t)\delta({\bf u})\delta({\bf x})
.\quad \end{eqnarray}

In the integral on the right hand side we perform a partial integration
 with respect to
${\bf z}$ and $s$. Since $f_1({\bf u},{\bf z},s)$ vanishes for 
$|{\bf z}|\rightarrow \infty$ the following boundary terms are obtained:
\begin{equation}\label{bound}
\left[
_0D_t^{1-\alpha} \int_{-\infty}^{\infty} d{\bf z} h({\bf z},{\bf x},s,t)
f_1({\bf u},{\bf z},s) \right]_{s=0}^{s=\infty}
\end{equation}
We can assume $h({\bf z},{\bf x};s=\infty,t)=0$.

Let us consider the boundary term using the initial condition 
$f_1({\bf u},{\bf z},s=0)=\delta({\bf u})\delta({\bf z})$.
In that case, the boundary term (\ref{bound}) yields:
\begin{eqnarray}
\left[_0D_t^{1-\alpha} \int_{-\infty}^{\infty} d{\bf z} h({\bf z},{\bf x};s,t)
f_1({\bf u},{\bf z},s) \right]_{s=0}^{s=\infty}=
-_0D_t^{1-\alpha} h(0,{\bf x};0,t)\delta({\bf u})=
-\delta({\bf x}) \delta({\bf u})\delta(t) \qquad .
\end{eqnarray}
This term cancels the last term on the right hand side of 
(\ref{7-Ob_FPEsubs}) provided we postulate the validity of the
boundary condition
\begin{equation}\label{bon}
_0D_t^{1-\alpha}h({\bf z}=0,{\bf x};s=0,t)=\delta(t)\delta({\bf x}) 
\qquad .
\end{equation}

As a consequence, 
the function $h({\bf z},{\bf x};s,t)$ has to obey the following
equation 
\begin{eqnarray}\label{6-PDE_H}
\left(\frac{\partial}{\partial t}
+{\bf u}\cdot\nabla_{{\bf x}}\right)h({\bf z},{\bf
x};s,t) &=& \:-\: _0D_t^{1-\alpha}\left(\frac{\partial}{\partial
s}+{\bf u}\cdot\nabla_{\bf z}\right)h({\bf z},{\bf x};s,t)
 \qquad,
\nonumber \\
_0D_t^{1-\alpha}h({\bf z}=0,{\bf x};s=0,t)&=& \delta(t)\delta({\bf x})
\qquad . 
\end{eqnarray}
This is the generalization of eq. (\ref{hdefine}).

We have to add the following remarks. First, 
equation (\ref{6-PDE_H}) determines the ${\bf u}$-independent
function $h({\bf z},{\bf x};s,t)$ by a relation including the velocity 
${\bf u}$. The fact that there is no ${\bf u}$-dependence
has been crucial for arriving at the equations 
(\ref{6-PDE_H}). Below, we shall show that we actually 
can find a ${\bf u}$-independent solution.
Second, we point out that the properties of the diffusion operator 
$L({\bf u},\nabla_u)$ do not show up in the determination of the function
$h({\bf z},{\bf x};s,t)$. The properties of this operator 
are included in the pdf $f({\bf u},{\bf z},s)$ and we have obtained
the solution in terms of an integral transform for a large class of
stochastic processes. 

\subsection{Determination of $h({\bf z},{\bf x};s,t)$}
   
The solution of differential equations containing the Riemann-Liouville 
fractional differential
operator is simplified by changing to Laplace-space. This is due to the
fact that the integral in the definition eq.(\ref{FracDiff_def}) is actually a
Laplace-convolution. Throughout this discussion we denote 
Laplace-transforms as
follows:
$\tilde{F}(\lambda):=\mathcal{L}\{F(t)\}:=\int_0^\infty dt\:e^{-\lambda
t}\:F(t)$. By performing a Laplace-transformation of eq.(\ref{6-PDE_H}) we
derive the following first order partial differential equation:
\begin{eqnarray}\label{lak}
\left(\lambda+{\bf u}\cdot\nabla_{{\bf x}}\right)\tilde{h}({\bf z},{\bf x};s,
\lambda)-h({\bf z},{\bf x};s,0)
=&-&\lambda^{1-\alpha}\left(\frac{\partial}{\partial s}+{\bf u}\cdot\nabla_{\bf
z}\right)\tilde{h}({\bf z},{\bf x};s,\lambda) \nonumber\\
&+&\left(\frac{\partial}{\partial s}+{\bf u}\cdot\nabla_{\bf
z}\right)\left[_0D_t^{1-\alpha}h({\bf z},{\bf x};s,t) \right]_{t=0}
\qquad.\end{eqnarray}
The last term on the right hand side is due to the partial time derivative in
the fractional differential operator. It can be set to zero:
$\left[_0D_t^{-\alpha}h({\bf z},{\bf x};s,t)\right]_{t=0}=0$. Equation 
(\ref{lak}) is valid for $s>0$. For $s=0$ we have the boundary condition
$h({\bf z},{\bf x};s,0)=\delta({\bf z}-{\bf x})\delta(s)$.  
Thus we obtain the final form of the
evolution equation for $\tilde{h}({\bf z},{\bf x};s,\lambda)$:
\begin{eqnarray} \label{7-Ob_PDEchar1}
\left\{\frac{\partial}{\partial
s}+{\bf u}\cdot\nabla_{{\bf z}}
+\lambda^{\alpha-1}{\bf u}\cdot\nabla_{{\bf x}}\right\}
\tilde{h}({\bf z},{\bf x};s,\lambda) =-\lambda^\alpha
\tilde{h}({\bf z},{\bf x};s,\lambda) \qquad.\end{eqnarray}

Linear partial differential equations of this type can be solved by 
the method of characteristics (see e.g. \cite{John}). 
This method will be applied below.\\
If we choose $s$ as parameter of the characteristics, eq.(\ref{7-Ob_PDEchar1})
can be written as
\begin{eqnarray}\label{7-Ob_PDEchar2}
\frac{d}{ds}\tilde{h}({\bf z}(s),{\bf x}(s);\lambda,s)&=&\left\{\frac{\partial}
{\partial s}
+\frac{d{\bf z}(s)}{ds}\cdot\nabla_{{\bf z}}+\frac{d{\bf
x}(s)}{ds}\cdot\nabla_{{\bf x}} \right\}\tilde{h}({\bf z}(s),{\bf x}(s);\lambda,s)
\nonumber \\ &=&-\lambda^\alpha\:\tilde{h}({\bf z}(s),{\bf x}(s);\lambda,s)
\qquad.\end{eqnarray}
Clearly the solution of this differential equation reads
\begin{eqnarray}\label{7-Ob_PDElsg}
\tilde{h}({\bf z}(s),{\bf x}(s);\lambda,s)=\tilde{h}({\bf
z}(0),{\bf x}(0);\lambda,0)\:e ^{-\lambda^\alpha s} \qquad.\end{eqnarray}

The characteristics ${\bf z}(s)$, ${\bf x}(s)$ are determined by
ordinary first order differential equations:
\begin{eqnarray}
\label{7-Ob_ODEchar}
\frac{d{\bf z}(s)}{ds}&=&{\bf u} \qquad\qquad\rightarrow\qquad
{\bf z}(s)={\bf u}s+{\bf z}_0 \qquad, \nonumber \\
\frac{d{\bf x}(s)}{ds}&=&\lambda^{\alpha-1}{\bf u} \qquad\rightarrow\qquad
{\bf x}(s)=\lambda^{\alpha-1}{\bf u}s+{\bf x_0} \qquad.\end{eqnarray}

For the initial condition $\tilde{h}({\bf z}(0),{\bf x}(0);0,\lambda)=\tilde{h}({\bf
z}_0,{\bf x}_0;0,\lambda)$ we specify a sharp distribution at ${\bf x}_0$. The
dependence on $\lambda$ is chosen as $\lambda^{\alpha-1}$ to be consistent
with the boundary condition (\ref{6-PDE_H}). If we
put ${\bf z}_0=0$ the inital condition takes the form:
\begin{eqnarray}\label{7-Ob_start}
\tilde{h}({\bf z}_0=0
,{\bf x}_0;0,\lambda)&=&\delta({\bf x}_0)\lambda^{\alpha-1}
=\delta({\bf x}-\lambda^{\alpha-1}{\bf z})\lambda^{\alpha-1}
\qquad.
\end{eqnarray}
Finally we obtain the kernel of the transformation eq.(\ref{7-Ob_LSG}) as
solution of the partial differential equation (\ref{7-Ob_PDEchar1}) in
Laplace-space:
\begin{eqnarray}\label{7-Ob_Hlsg}
\tilde{h}({\bf z},{\bf x};s,\lambda)
&=&\delta({\bf x}-\lambda^{\alpha-1}{\bf z})
\lambda^{\alpha-1}e^{-\lambda^\alpha s}=\delta({\bf x}-\lambda^{\alpha-1}{\bf
z})\tilde{h}_\alpha(s,\lambda) \qquad.
\end{eqnarray}
In the last step we substitute the Laplace-transform of $h_\alpha(s,t)$:
$\mathcal{L}\{h_\alpha(s,t)\}=\lambda^{\alpha-1}e^{-\lambda^\alpha s}$.\\
From this expression the integral kernel in physical time $h({\bf z},{\bf
x};s,t)$ is given as inverse Laplace-transform:
\begin{eqnarray}\label{inv}
h({\bf z},{\bf x};s,t)
=\mathcal{L}^{-1}\{\delta({\bf x}-\lambda^{\alpha-1}{\bf z})\}*h_\alpha(s,t)
\qquad.\end{eqnarray}
The asterisk $*$ denotes a Laplace-convolution with respect to $t$. However due
to the $\lambda$-dependence of the $\delta$-function a closed form of the
inverse transformation $\mathcal{L}^{-1}\{\delta({\bf x}-\lambda^{\alpha-1}{\bf
z})\}$ could not be calculated and the explicit expression of $h({\bf z},{\bf
x};s,t)$ remains unknown.\\
Still the transformation eq.(\ref{7-Ob_LSG}) provides a useful equation
for characterizing the solution $f({\bf u},{\bf x},t)$. Since inverse
Laplace-transform and integration commute we can give an expression for
the transformed solution $\tilde{f}({\bf
u},{\bf x},\lambda):=\mathcal{L}\{f({\bf u},{\bf x},t)\}$:
\begin{eqnarray}\label{7-Ob_LSG_L} \tilde{f}({\bf u},{\bf
x},\lambda)=\int_{-\infty}^\infty d{\bf z} \int_0^\infty ds\:\tilde{h}({\bf
z},{\bf x};s,\lambda)f_1({\bf u},{\bf z},s) \qquad.\end{eqnarray}

Thus statistical quantities can be calculated in Laplace-space and then
transformed to physical time. It will become clear in the following
that with the help of the integral kernel 
$\tilde{h}({\bf z},{\bf x};s,\lambda)$
the solution of the generalized Obukhov-model can be completely characterized.

However, several remarks are in order.
First of all, one can easily check that the solutions
$f({\bf u},{\bf x},t)$ of eq.(\ref{model}) are normalized:
\begin{eqnarray}\label{norma}
\int_{-\infty}^\infty d{\bf x} \int_{-\infty}^\infty
d{\bf u}\:f({\bf u},{\bf x},t)=\int_{-\infty}^\infty d{\bf u}\:f({\bf u},t)=1
\qquad.\end{eqnarray}
Additionally one states that $\tilde{h}({\bf z},{\bf x};s,\lambda)$ 
is always nonegative and
so are $h({\bf z},{\bf x};s,t)$, $f_1({\bf u},{\bf z},s)$, and,
in turn, $f({\bf u},{\bf x},t)$. This proofs that the generalized Oboukhov 
model actually defines a probability distribution for the case 
$0<\alpha<1$.
\\

Second, we have to show that the above assumptions are fullfilled.
To this end we note that the quantity $h({\bf z},{\bf x};s,t)$ in fact 
is independent on the velocity
${\bf u}$. Furthermore, we have the relationship
\begin{equation}
\tilde{h}(0,{\bf x};s,\lambda)=\delta({\bf x})\tilde{h}_\alpha(s,\lambda)
\qquad ,
\end{equation}
which in real space reads
\begin{equation}
_0D_t^{1-\alpha} h(0,{\bf x};0,t)=
\delta({\bf x}) _0D_t^{1-\alpha} h_\alpha(0,t)=
\delta({\bf x}) \qquad .
\end{equation}
Since $h_\alpha(s,0)=\delta(s)$ we obtain 
\begin{equation}
h({\bf z},{\bf x};s,0)=\mathcal{L}^{-1}\{\delta({\bf
x}-\lambda^{\alpha-1}{\bf z})\}*\delta(s)\sim \delta(s) \qquad .
\end{equation}
\\
The case $\alpha=1$ has to lead to the ordinary Obukhov model.
Evidently the integral kernel eq.(\ref{7-Ob_Hlsg}) in 
this limit is given by the Laplace transform
\begin{eqnarray}\label{lk}
\tilde{h}({\bf z},{\bf x};s,\lambda)=\delta({\bf x}-{\bf z})e^{-\lambda s}
\qquad .
\end{eqnarray}
The inverse Laplace-transformation is easily calculated:
\begin{eqnarray}\label{lk2}
h({\bf z},{\bf x};s,t)=\delta({\bf x}-{\bf z})\delta(t-s)
\qquad,\end{eqnarray}
such that the transformation eq.(\ref{7-Ob_LSG}) leads to the solution
of the ordinary Obukhov-model.\\
\\
The second limit case is supposed to yield the solution $G({\bf u},t)$
of the fractional diffusion equation. Performing the integration of
eq.(\ref{7-Ob_LSG_L}) with respect to ${\bf x}$ results in:
\begin{eqnarray}
\int_{-\infty}^\infty d{\bf x}\int_{-\infty}^\infty d{\bf z} \int_0^\infty ds\:
\delta({\bf x}-\lambda^{\alpha-1}{\bf z})
\tilde{h}_\alpha(s,\lambda) f_1({\bf u},{\bf
z},s)=\int_0^\infty ds\:\tilde{h}_\alpha(s,\lambda) G_1({\bf u},s)
\quad. \end{eqnarray}
The right hand side of this equation is the Laplace-transform of the
solution eq.(\ref{FDE_sol}) of the fractional diffusion equation
(\ref{FDE}). The agreement in this case is achieved due to 
the choice $\lambda^{\alpha-1}$
in the initial condition eq.(\ref{7-Ob_start}).

\subsection{The joint probability distribution $f({\bf u},{\bf x},t)$}

In the following we derive a formal expression for the joint 
probability distribution
$\tilde{h}({\bf z},{\bf x};s,\lambda)$. 
First the Laplace-transform of $f({\bf u},{\bf x},t)$
is calculated using the transformation eq.(\ref{7-Ob_LSG_L}):
\begin{eqnarray}\label{pdflap}
f({\bf u},{\bf x},\lambda)&=&\int_{-\infty}^\infty d{\bf z} \int_0^\infty ds\:
\delta({\bf x}-\lambda^{\alpha-1}{\bf z})\tilde{h}_\alpha(s,\lambda)
f_1({\bf u},{\bf z},s) \nonumber\\
&=&\int_0^\infty
ds\:\lambda^{d(1-\alpha)}\tilde{h}_\alpha(s,\lambda)
f_1({\bf u},\lambda^{1-\alpha}{\bf x},s)
\qquad. \end{eqnarray}
Here the integration with respect to ${\bf z}$ is performed in 
$d$ dimensions. Let us assume that the diffusion operator $L({\bf u},\nabla_u)$
allows for solutions exhibiting scaling behaviour of the form
\begin{equation}\label{scal} 
f_1({\bf u},{\bf z},s)= \frac{1}{s^{(1+2\delta)d}}
H(\frac{{\bf z}}{s^{1+\delta}},\frac{{\bf u}}{s^{\delta}})
\end{equation}

This yields:
\begin{equation}
f({\bf u},{\bf x},\lambda)
=\int_0^\infty
ds\:\lambda^{d(1-\alpha)}\tilde{h}_\alpha(s,\lambda)
\frac{1}{s^{(1+2\delta)d}}
H(\frac{\lambda^{1-\alpha} {\bf x}}{s^{1+\delta}},\frac{{\bf u}}{s^{\delta}})
\qquad. 
\end{equation}

Let us consider the case of the simple diffusion operator 
$L=\Delta$. $f_1({\bf u},{\bf z},s)$
is then explicitly given as:
\begin{eqnarray}
\label{7-Ob_Gauss}
f_1({\bf u},{\bf z},s)=\left( \frac{\sqrt{3}}{2\pi s^2} \right)^d
\:\exp\left\{-\frac{{\bf u}^2}{s}+\frac{3{\bf z}
\cdot {\bf u}}{s^2}-\frac{3{\bf z}^2}{s^3}\right\} \qquad.
\end{eqnarray}
This pdf exhibits scaling behaviour of the form (\ref{scal})
with $\delta=1/2$.

We have not determined the Laplace inversion of this formula. However,
we shall obtain the probability distributions for the velocity ${\bf u}$
and the spatial variable ${\bf x}$.

\subsection{The probability distribution $G({\bf x},t)$}

We shall start with the probability distribution $G({\bf u},t)$ for the
velocity ${\bf u}$ which is obtained by an integration with respect to the 
spatial variable ${\bf x}$:
\begin{equation}
G({\bf u},t)= \int d{\bf x} f({\bf u},{\bf x},t) \qquad .
\end{equation}
In Laplace space, we obtain
\begin{equation}
\tilde G({\bf u},\lambda)= \int_0^\infty 
ds \lambda^{\alpha-1} e^{-\lambda^\alpha s}
G_1({\bf u},s) \qquad ,
\end{equation}
which, by 
Laplace inversion, yields
\begin{equation}
G({\bf u},t)= \int_0^\infty ds h_\alpha(s,t) G_1({\bf u},s) \qquad .
\end{equation}
If we assume that the function $G_1({\bf u},s)$ exhibits scaling
behaviour,
\begin{equation}
G_1({\bf u},s)=\frac{1}{s^{d\delta}}g(\frac{{\bf u}}{s^\delta})
\qquad ,
\end{equation}
the result may be reexpressed in a form which just evidences this
scaling behaviour
\begin{equation}\label{fg}
G({\bf u},t)= \int d\sigma h_\alpha(\sigma)\frac{1}{t^{d \delta\alpha}}
g(\frac{{\bf u}}{t^{\delta \alpha}}\sigma^{\alpha \delta})
= \frac{1}{t^{d\delta \alpha}} \tilde g(\frac{{\bf u}}{t^{\delta\alpha}})
\qquad .
\end{equation}
 
Let us consider the case for $L=\Delta_u$. Here,
$\delta=\frac{1}{2}$:
\begin{equation}
G({\bf u},t)= \int ds \frac{1}{\alpha} \frac{t}{s^{1+1/\alpha}}
L_\alpha (\frac{t}{s^{1/\alpha}})
\frac{1}{(\sqrt{4 \pi s})^d}e^{-\frac{{\bf u}^2}{4s}} \qquad .
\end{equation}
This may also be represented according to
\begin{equation}\label{vel}
G({\bf u},t)= \int d\sigma 
L_\alpha (\sigma)
\frac{\sigma^{d\alpha/2}}{(\sqrt{4 \pi t^\alpha})^3}
e^{-\frac{{\bf u}^2}{4t^\alpha} \sigma^\alpha} \qquad .
\end{equation}

Formula (\ref{fg}) yields an interesting structure. 
First of all, it represents 
a probability distribution as a superposition of pdf's $G_1({\bf x},s)$ 
with different variances. Such a representation has been introduced by
Castain et al. \cite{Castain} in a description of the intermittent behaviour
of the Eulerian velocity increment pdf. Second, scaling behaviour 
$<{\bf u}^2(t)>\sim t$ is quite often related to a random walk
behaviour in velocity space. However, the formula (\ref{fg}) shows that
such scaling behaviour may also arise for the cases \cite{Friedrich}
where the scaling indices
$\alpha$, $\delta$ are related by 
\begin{equation}
\delta \alpha =\frac{1}{2} \qquad .
\end{equation}
This e.g. happens for the rotationally symmetric solutions 
of the diffusion operator 
\begin{equation}
L({\bf u},\nabla_u)= \nabla ({\bf u}^2)^{(1-\alpha)} \nabla \qquad .
\end{equation}
Here, the pdf $G({\bf u},s)$ is given by
\begin{equation}
G({\bf u},s)= N(\alpha) \frac{1}{s^{\frac{d}{2\alpha}}}
e^{-\frac{{\bf u}^{2\alpha}}{4\alpha^2 s}}
\end{equation}
These solutions have been shown to yield accurate reconstructions
of the velocity increment pdf for turbulent flows measured in
\cite{Pinton} by adjusting the
parameter $\alpha$ \cite{Friedrich}. For the case of
turbulent flows, the determination of the parameter
$\alpha$ and especially its time dependence is an open
problem. 

\subsection{The probability distribution $F({\bf x},t)$}

Let us now turn to the probability distribution 
$F({\bf x},t)$ for the position ${\bf x}$. This distribution is obtained
from the joint pdf $f({\bf u},{\bf x},t)$ by integration with respect to 
the velocity variable:
\begin{equation}
F({\bf x},t)= \int d{\bf u} f({\bf u},{\bf x},t) \qquad .
\end{equation}
The Laplace transform of this quantity can be determined from 
(\ref{pdflap})
\begin{equation}
F({\bf x},\lambda)=
\int_0^\infty  ds \lambda^{d(1-\alpha)} h_\alpha(s,\lambda) 
F_1(\lambda^{1-\alpha}{\bf x},s) \qquad .
\end{equation} 
If we consider a system exhibiting scaling behaviour,
\begin{equation}
F_1({\bf z},s)=\frac{1}{s^{d(1+\delta)}}F(\frac{{\bf z}}{s^{(1+\delta)}})
\qquad ,
\end{equation}
we may perform the transformation
\begin{equation}
S=s \lambda^{-(1-\alpha)/(1+\delta)} 
\end{equation}
leading to 
\begin{equation}\label{dec}
F({\bf x},\lambda)=\int_0^\infty dS \lambda^{(1-\alpha)/(1+\delta)}
h_\alpha
(S \lambda^{(1-\alpha)/(1+\delta)},\lambda) \frac{1}{S^{d(1+\delta)}}
F(\frac{{\bf x}}{S^{(1+\delta)}}) \qquad .
\end{equation}
Laplace inversion yields the probability distribution 
\begin{equation}
F({\bf x},t)=\int_0^\infty dS \frac{1}{S^{d(1+\delta)}}
F(\frac{{\bf x}}{S^{(1+\delta)}})
h_{\frac{\delta \alpha+1}{\delta+1}}(s,t) \qquad .
\end{equation}
The decisive step thereby, is the possibility to calculate 
the Laplace inverse of the function
\begin{eqnarray}
\mathcal{L}^{-1} \lbrace \lambda^{(1-\alpha)/(1+\delta)}
h_\alpha(S \lambda^{(1-\alpha)/(1+\delta)},\lambda) \rbrace
&=&
\mathcal{L}^{-1} 
\lbrace
\lambda^{\frac{\alpha\delta+1}{1+\delta}-1} 
e^ {-\frac{\alpha \delta+1}{1+\delta}S}
\rbrace
\nonumber \\
&=& \frac{\delta+1}{\alpha\delta+1}
\frac{t}{s^{1+\frac{1+\delta}{\alpha \delta+1}}}
 L_{\frac{\alpha \delta+1}{1+\delta}}(\frac{t}
{s^{\frac{1+\delta}{\alpha \delta+1}}})
=h_{\frac{\delta \alpha+1}{\delta+1}}(s,t) \qquad .
\end{eqnarray}

We can now explicitly state the result for the diffusion operator
$L=\Delta_u$:
\begin{equation}
F({\bf x},t)= \int_0^\infty dS e^{-\frac{3 {\bf x}^2}{4S^3}}
\left( \sqrt{\frac{3}{4 \pi S^3}} \right)^d \frac{3}{2+\alpha}
L_{(2+\alpha)/3} \left( \frac{t}{S^{3/(2+\alpha)}} \right)
\frac{t}{S^{1+3/(2+\alpha)}} \qquad .
\end{equation}
One observes that the limiting case $\alpha=1$ leads to pdf of the
Obukhov model. Furthermore, it is evident that
the probability distribution of the
spatial variable ${\bf x}$ is given by an expression similar to the one 
obtained for the velocity variable in form of an integral transform:
\begin{equation}
F({\bf x},t)= \int d\sigma 
L_{(2+\alpha)/3} (\sigma)
(\sqrt{\frac{3}{4 \pi t^{2+\alpha}}})^d \sigma^{3(2+\alpha)/2}
e^{-\frac{3{\bf x}^2}{4t^{2+\alpha}} \sigma^{2+\alpha}}
\qquad .
\end{equation}
This expression has to be compared with the one obtained for the 
pdf of the velocity (\ref{vel}). 

We mention that the solution (\ref{dec}) 
can be related to a continuous time random
walk for the position increment ${\bf x}(t)$ of a Lagrangian particle.
Continuous time random walks for the description of 
the distance between two Lagrangian variables in connection
with Richardson's law have been discussed
by Shlesinger et al. \cite{Shlesinger}.

\subsection{Moments of the generalized Obukhov-model}

Starting from eq.(\ref{pdflap}) we can derive a general relation 
for moments of arbitrary order of the joint probability distribution
$f({\bf u},{\bf x},t)$.\\
Single-time moments $<x_i^n(t)u_i^m(t)>$ are defined as follows:
\begin{eqnarray}\label{6-EOb_moment1}
<x_i^n(t)u_i^m(t)>\:=\int_{-\infty}^\infty d{\bf u}
 \int_{-\infty}^\infty
d{\bf x} \: x_i^n u_i^m\:f({\bf u},{\bf x},t)
\qquad.\end{eqnarray}
Therefore we obtain in Laplace-space with the help of 
eq.(\ref{7-Ob_LSG_L}) and eq.(\ref{7-Ob_Hlsg}):
\begin{eqnarray}
\label{6-EOb_moment2}
\mathcal{L}\{<x_i^n(t)u_i^m(t)>\} &=&\int_{-\infty}^\infty d{\bf u} \int_{-\infty}^\infty d{\bf z} \int_0^\infty
ds\:
\delta({\bf x}-\lambda^{\alpha-1}{\bf z})\tilde{h}_\alpha(s,\lambda)x_i^n 
u_i^m\: f_1({\bf u},{\bf z},s)\nonumber\\
&=&\lambda^{(\alpha-1)n}\int_0^\infty ds\:\tilde{h}_\alpha(s,\lambda)
\int_{-\infty}^\infty d{\bf u} \int_{-\infty}^\infty
d{\bf z}\: z_i^n
u_i^m\:f_1({\bf u},{\bf z},s) \qquad. \end{eqnarray}
In the last step the integration over the $\delta$-function 
has been performed. Here we recognize that
simply the moments of the ordinary Obukhov-model appear:
$\int d{\bf u} \int
d{\bf z}\: u_i^m z_i^n\:f_1({\bf u},{\bf z},s)
=\:<z_i^n(s)u_i^m(s)>$. As as consequence
the inverse Laplace-transform leads to a general 
expression for the moments of the generalized model:
\begin{eqnarray}\label{6-EOb_moment_allg}
<x_i^n(t)u_i^m(t)>\:&=&\:_0D_t^{(\alpha-1)n}\int_0^\infty
ds\:h_\alpha(s,t)\:<z_i^n(s)u_i^m(s)> \qquad.\end{eqnarray} 
Thus the moments $<x_i^n(t)u_i^m(t)>$ can be determined 
from the moments of the ordinary model
$<z_i^n(s) u_i^m(s)>$ by an inverse L\'evy-transform and fractional
integration with respect to $t$. Since $<z_i^n(s)u_i^m(s)>$ 
take the form of simple polynomials
in $s$ this results in fractional integration of single-time moments of the
inverse L\'evy-stable process \cite{Baule}. These calculations can always
be performed in an elementary manner.\\ \\

For example, by applying eq.(\ref{6-EOb_moment_allg}) to the  
second order moments for the case of $L=\Delta_u$  we obtain:
\begin{eqnarray}
\label{6-EOb_moment_sec}
<{\bf u}^2(s)>&=&2 s\qquad\quad\rightarrow\qquad<{\bf u}^2(t)>\:=
\frac{2}{\Gamma(\alpha+1)}
t^\alpha\qquad,\nonumber\\ <{\bf z}^2(s)>&=&\frac{2}{3}s^3
\qquad\rightarrow\qquad<{\bf x}^2(t)>
\:=\frac{4}{\Gamma(\alpha+3)}t^{\alpha+2}
\qquad, \nonumber\\ <{\bf z}(s){\bf u}(s)>&=&s^2 
\qquad \rightarrow
\qquad <{\bf x}(t){\bf u}(t)>
\:=\frac{2}{\Gamma(\alpha+2)}t^{\alpha+1}.
\end{eqnarray}
Clearly, in each case the limit $\alpha \rightarrow 1$ 
of the Obukhov-model is
satisfied. Therefore the generalized Obukhov-model eq.(\ref{model}) 
with $L=\Delta_u$ is
characterized by subdiffusive behaviour of the particle velocity 
corresponding to the fractional diffusion equation (\ref{FDE}) and mean
square displacement $<{\bf x}^2(t)>\:\sim t^{\alpha+2}$ of the particle
position. 

\section{Conclusion and outlook}

We have introduced a non-Markovian model for the joint position-velocity
probability distribution of a random walk particle as a generalization of
Obukhov's model. Long-time correlations of the particle velocity are
taken into account by fractional time derivatives. As in the limit case
of the fractional diffusion equation, a solution
can be found as an integral transformation of the Markovian solution.
The integral kernel has been obtained by solving a fractional differential
equation and yields a complete characterization of the single time
statistical properties. 
As final results we have given analytical expressions for the joint
probability distribution and its moments. The
generalized Obukhov-model may thus serve as an appropriate model for
random walk processes revealing anomalous diffusion in velocity- and
position-space.

\acknowledgements
We gratefully acknowledge support by the Deutsche
Forschungsgemeinschaft and wish to thank
R. Hillerbrand, O. Kamps and T. D. Frank for helpful discussions.

\end{document}